\documentclass[useAMS,usenatbib]{mn2e}
\usepackage{amsmath,amssymb}
\usepackage{graphicx}
\title[stochastic backgrounds of gravitational waves]{Stochastic backgrounds of gravitational waves from cosmological sources -
The role of dark energy}
\author[O. D. Miranda]{Oswaldo D. Miranda\thanks{E-mail: oswaldo@das.inpe.br}\\
INPE - Instituto Nacional de Pesquisas Espaciais - Divis\~{a}o de Astrof\'{i}sica,\\
Av. dos Astronautas 1758, S\~{a}o Jos\'{e} dos Campos, 12227-010 SP, Brazil}
\begin{document}

\date{Accepted  . Received ; in original form }

\pagerange{\pageref{firstpage}--\pageref{lastpage}} \pubyear{2011}

\maketitle

\label{firstpage}

\begin{abstract}
In this work we investigate the detectability of the gravitational stochastic background produced by cosmological sources
in scenarios of structure formation. The calculation is performed in the framework of hierarchical structure formation
using a Press-Schechter-like formalism. The model considers the coalescences of three kind of binary systems, namely
double neutron stars (NS-NS), the neutron star-black hole binaries (NS-BH), and the black hole-black hole systems
(BH-BH). We also included in the model the core-collapse supernovae leaving black holes as compact remnants. In
particular, we use two different dark-energy scenarios, specifically cosmological constant ($\Lambda$) and Chaplygin gas,
in order to verify their influence on the cosmic star formation rate, the coalescence rates, and on the  gravitational
wave backgrounds. We calculate the gravitational wave signals separately for each kind of source as well as we determine
their collective contribution for the stochastic background of gravitational waves. Concerning to the compact binary
systems, we verify that these sources produce stochastic backgrounds with signal-to-noise ratios (S/N) $\sim 1.5$
($\sim 0.90$) for NS-NS, $\sim 0.50$ ($\sim 0.30$) for NS-BH, $\sim 0.20$ ($\sim 0.10$) for BH-BH, and $\sim 0.14$
($\sim 0.07$) for core-collapse supernovae for a pair of advanced LIGO detectors in the cosmological constant
(Chaplygin gas) cosmology. Particularly, the sensitivity of the future third generation of detectors as, for example, the
Einstein Telescope (ET), in the triangular configuration, could increase the present signal-to-noise ratios by a
high-factor ($\sim 300 - 1000$) when compared to the (S/N) calculated for advanced LIGO detectors. As an example, the
collective contribution of these sources can produce $({\rm S/N})\sim 3.3$ ($\sim 1.8$) for the $\Lambda$ (Chaplygin gas)
cosmology for a pair of advanced LIGO interferometers and within the frequency range $\sim 10\,{\rm Hz} 
- 1.5\, {\rm kHz}$. Considering ET we have $({\rm S/N})\sim 2200$ ($\sim 1300$) for the $\Lambda$ (Chaplygin gas)
cosmology. Thus, the third generation of gravitational wave detectors could be used to reconstruct the history of star
formation in the Universe as well as for contributing with the characterization of the dark energy, for example,
identifying if there is evidence for the evolution of the dark energy equation-of-state parameter $w(a)$.
\end{abstract}


\section[]{Introduction}

The direct detection of gravitational waves (GWs) is a major challenge for the physics/astrophysics and considerable experimental
effort is being devoted by several groups around the world. In particular, the window in the frequency range $\sim 10-10\;{\rm kHz}$
is open due to pioneerism of the following interferometers: LIGO detectors (e.g., \citealt{abbott09}), VIRGO detector
(e.g. \citealt{acernese08}), GEO 600 detector (e.g., \citealp{grote08}), and TAMA 300 detector (e.g., \citealt{tak04}).

In the future, another window in the low frequency range $\sim 10^{-4} -10\;{\rm Hz}$ will be open by space antennas such
as e-LISA/NGO \citep{elisa}, BBO \citep{cuha06}, DECIGO \citep{ando10}. These interferometers together ground-based detectors
which are presently in project, as KAGRA \citep{kagra}, and the third generation of resonant mass
detectors as SCHENBERG (\citealt{aguiar08}) and MiniGRAIL (\citealt{gott07}) will transform the research in general
relativity into an observational/theoretical study.

On the other hand, as GWs are produced by a large variety of astrophysical sources and cosmological phenomena,
it is quite probable that the Universe is pervaded by a background of such waves. Collapse of Population II and III stars,
phase transitions in the early Universe, cosmic strings, and a variety of binary stars are some examples of sources that
could produce such a putative background of GWs (see, e.g., \citealt{mag1,regimbau01,d3,d4,d5,d2,m1,sand1,suwa,gio,pereira10} among others).

Observe that the indirect evidence for the existence of gravitational waves came first from observations of the orbital decay
of the Hulse-Taylor binary pulsar \citep{hulse1,hulse2,hulse3}. In this century, direct detection though and analysis of
gravitational-wave sources are expected to provide a unique insight to one of the least understood of the fundamental forces
\citep{belcz}. Specifically, gravitational waves could also be used as a tool for studying the viability of different alternative
theories of gravity. This could be done by comparing the detected polarization modes with those predicted by general relativity
(see, e.g., \citealt{wayne,alves}).

As mentioned above, a number of interferometers designed for gravitational wave detection are currently in operation,
being developed, or planned. In particular, the high frequency part of the gravitational wave spectrum ($10\,{\rm Hz} 
\lesssim f \lesssim 10\,{\rm kHz}$) is open today through the pioneering efforts of the first-generation ground-based
interferometers such as LIGO. While detections from this first generation of detectors are likely to be rare,
the third generation of gravitational wave detectors as, for example, the Einstein Telescope (ET) may detect, among
others, the stochastic signal generated by a population of pre-galactic stars. Thus, gravitational wave observations
could add a new dimension to our ability to observe and understand the Universe.

On the other hand, the state of the art in cosmology has led to the following distribution of the energy densities of the Universe:
$4\%$ for baryonic matter, $23 \%$ for non-baryonic dark matter and $73 \%$ for the so-called dark energy (\citealp{j2011}). Concerning to the dark energy,
some equations of state have been proposed in order to explain such a dark component. The most common example is the cosmological constant ($\Lambda$CDM model),
which implies on a constant vacuum energy density along the history of the Universe. Another possibility is a dynamical vacuum or quintessence. In general, the
quintessence models involve one (\citealp{alb2000}) or two (\citealp{bento}) coupled scalar fields. The Chaplygin gas is another example of dark energy fluid.
One of the most appealing aspects of the original Chaplygin gas model is that it is equivalent to the Dirac-Born-Infeld (DBI) description of a Nambu-Goto membrane
\citep{bord94,gor,oga10}.

In this way, the main goal of the present work is to explore the possibility of using stochastic backgrounds of gravitational waves to provide more
information about the character and interrelationship of the dark-energy equation of state with the star formation at high redshifts. That is, in first place we
analyze the influence of two different dark-energy components of the Universe, namely cosmological constant and Chaplygin gas, on the stochastic backgrounds
of gravitational waves produced by four different cosmological sources which are the merging together of two neutron stars (NS-NS), the coalescence
of neutron star-black hole (NS-BH) systems, the merger of two black holes (BH-BH), and the core-collapse supernovae leaving
black holes as compact remnants. In second place, we show that different dark-energy fluids produce distinct signatures for
the cosmic star formation rate (CSFR) specially at high redshifts ($z > 3$). This interesting feature could be used as an
alternative way to study the star formation and the possible temporal dependence of the dark-energy equation of state up
to redshift 20, having as the common tool a stochastic background of gravitational waves with high signal-to-noise ratio.
In this way, not only binary systems at lower redshifts ($z < 2-3$) working as standard sirens but also stochastic backgrounds
of gravitational waves could contribute for a better comprehension of the physical nature of the dark energy and their
connection, and influence, with the star formation at high redshifts. 

The preference for concentrating attention on the Chaplygin gas also comes from recent work of \cite{pace10} who analyzed the spherical collapse model in dark-energy cosmologies. As can be seen from that work, the Chaplygin gas exhibits an equation-of-state dependent on time as the quintessence models also exhibit. It is not the purpose of the present study make an individual assessment of each particular type of dark-energy candidate. Our goal is to verify if stochastic backgrounds of gravitational waves can give us some indication about the evolution of the dark-energy equation of state with time. Thus, the comparison between Chaplygin gas and $\Lambda$-CDM is sufficient for the purposes of this study.

Here, we start with the CSFR recently derived by \citet{pereira10}. Specifically, these authors use a hierarchical structure
formation model and they obtain the CSFR in a self-consistent way. That means, the authors solve the equation governing the total
gas density taking into account the baryon accretion rate, treated as an infall term, and the lifetime of the stars formed in
the dark matter haloes. Here, we adapted the formalism derived by \cite{pereira10} in order to obtain the CSFR and the coalescence
rates in consistency with the assumed dark-energy model.

The paper is organized as follows. In Section 2, we review the basics of the hierarchical model and how to obtain the
CSFR up to redshift $z \sim 20$ as a function of the specific dark-energy cosmology. In Section 3, we discuss how to
obtain the coalescence rates for NS-NS, NS-BH, and BH-BH systems from the CSFR. In Section 4, we present the formalism
used to characterize the gravitational wave backgrounds for compact binary systems and core-collapse supernovae to form
black holes. We also present the signal-to-noise ratios (S/N) for both a pair of advanced LIGO detectors and the ET in
triangular configuration. Section 5 presents the collective contribution of these sources for the gravitational
wave background. In Section 6 we discuss the influence of the uncertainties of the parameters on the derived
gravitational wave background and on the CSFR. Section 7 presents the final considerations of this work.

\section[]{the cosmic star formation rate and the dark energy cosmologies}

\subsection{An overview}

The essence of the halo model was discussed by \citet{neyman} who postulated that all galaxies form in clusters, the
distribution of galaxies within clusters can be described by a probabilistic relation, and cluster centers are themselves
correlated. Substituing the word `clusters' by `haloes' in the paper of \citet{neyman} we arrive at a reasonable
qualitative description of the modern halo model. Today, it is widely believed that haloes, or overdense dark matter
clumps, form as result of the growth and non-linear evolution of density perturbations produced in the early Universe
\citep{p1}.  This is the heart of the hierarchical formation scenario.

In general, the halo mass function is represented as the differential number density of haloes with mass between
$M$ and $M+dM$. Press and Schechter (hereafter PS) heuristically derived a mass function for bound virialized objects
in 1974 \citep{p2}. The basic idea of the PS approach is define haloes as concentrations of mass that have already left
the linear regime by crossing the threshold $\delta_{\rm c}$ for non-linear collapse. Given a power spectrum and a window
function, it should then be relatively straightforward to calculate the halo mass function as a function of the mass and
redshift. 

However, it is worth stressing that the exact definition of the mass function, e.g., integrated versus differential form
or count versus number density, varies widely in the literature. To characterize different fits, it can be introduced the
scale differential mass function $f(\sigma,z)$ \citep{j1} defined as a fraction of the total mass per
$\ln \sigma^{-1}$ that belongs to haloes. That is,

\begin{equation}
f(\sigma,z)\equiv\frac{d\rho/\rho_{\rm B}}{d\ln\sigma^{-1}}=\frac{M}{\rho_{\rm B}(z)}
\frac{dn(M,z)}{d\ln[\sigma^{-1}(M,z)]},
\end{equation}

\noindent where $n(M,z)$ is the number density of haloes with mass $M$, and $\rho_{\rm B}(z)$ is the background density at
redshift $z$. As pointed out by \citet{j1}, this definition of the mass function has the advantage that it does not
explicitly depend on redshift, power spectrum, or cosmology; all of these are contained in $\sigma(M,z)$ (see also
\citealt{luk}). See that

\begin{equation}
 \sigma(M,z)=\sigma(M,z=0)D(z),\label{sigmamz}
\end{equation}

\noindent is the linear rms density fluctuation in spheres of comoving radius $R$ containing the mass $M$, and $D(z)$
is the linear growth function.

The density of baryons is proportional to the density of dark matter if we consider that the baryon distribution traces
the dark matter. Thus, the fraction of baryons at redshift $z$ that are in structures is given by (see, e.g.,
\citealt{d1,pereira10})

\begin{equation}
f_{\rm b}(z)=\frac{\int_{M_{\rm min}}^{M_{\rm max}} {f(\sigma)MdM}}{\int_{0}^{\infty} {f(\sigma)MdM}}\label{fbaryon}
\end{equation}

\noindent where we have used $M_{\rm min}= 10^{6}\, {\rm M}_{\odot}$ and $M_{\rm max}= 10^{18}\, {\rm M}_{\odot}$ (see \citealp{pereira10} for details).
 
Therefore, the baryon accretion rate $a_{\rm b}(t)$ which accounts for the increase in the fraction of baryons in structures is given by

\begin{equation}
a_{\rm b}(t) = \Omega_{\rm b}\rho_{\rm c}\left(\frac{dt}{dz}\right)^{-1}\left|\frac{df_{\rm b}(z)}{dz}\right|,\label{abaryon}
\end{equation}

\noindent where $\rho_{\rm c}=3H_{0}^{2}/8\pi G$ is the critical density of the Universe.

The age of the Universe that appears in (\ref{abaryon}) is related to the redshift by:
 
\begin{equation}
\frac{dt}{dz} = \frac{9.78h^{-1} \rm{Gyr}}{(1+z)E(z)}.\label{timez}
\end{equation}

In Eq. (\ref{timez}) $E(z)$ represents the expansion function which is (see, e.g., \citealp{pace10})

\begin{equation}
 E(z) = \sqrt{\Omega_{\rm m}(1+z)^{3}+\Omega_{\rm d}\exp{\left(-3\int_{1}^{a}\frac{1+w(a')}{a'}da\right)}}\label{expan},
\end{equation}

\noindent where the relative density of the i-component is given by $\Omega_{\rm i}=\rho_{\rm i}/\rho_{\rm c}$, and `i'
appying for baryons (b), dark energy (d), and total matter (m). As usual, the scale factor is $a=1/(1+z)$, and $w(a)$ is
the dark energy equation-of-state parameter.

Note that for $w(a)= -1$ we have the equation-of-state parameter of the cosmological constant. In this case

\begin{equation}
 E(z) = \sqrt{\Omega_{\rm m}(1+z)^{3}+\Omega_{\Lambda}}\label{expan2},
\end{equation}

\noindent with $\Omega_{\rm d} = \Omega_{\Lambda}$.

The linear growth function, in Eq. (\ref{sigmamz}), is defined as $D(z) \equiv \delta_{\rm m}(z)/\delta_{\rm m}(z=0)$ and it is obtained
as a solution from the following equation (see \citealp{pace10} for details)

\begin{equation}
 \delta_{\rm m}''+\left(\frac{3}{a}+\frac{E'}{E}\right)\delta_{\rm m}' - \frac{3}{2} \frac{\Omega_{\rm m}}{a^{5}E^{2}}\delta_{\rm m}=0,
 \label{growth}
\end{equation}

\noindent where the derivatives are taken in relation to the scale factor $a$.

On the other hand, the equation governing the total gas mass ($\rho_{\rm g}$) in the haloes is

\begin{equation}
 \dot\rho_{\rm g}=-\frac{d^{2}M_{\star}}{dVdt}+\frac{d^{2}M_{\rm ej}}{dVdt}+a_{\rm b}(t)\label{rhogas}.
\end{equation}

The first term on the right hand side of equation (\ref{rhogas}) represents the stars which are formed by the gas contained in the haloes.
Using a Schmidt law \citep{sch1,sch2} we can write for the star formation rate 

\begin{equation}
\frac{d^{2}M_{\star}}{dVdt} = \Psi(t) = k\rho_{\rm g}(t),\label{sclaw}
\end{equation}

\noindent where $k$ is the inverse of the time-scale for star formation. That is, $k=1/\tau_{\rm s}$.

The second term on the right hand side of equation (\ref{rhogas}) considers the mass ejected from stars through winds and supernovae.
Therefore, this term represents the gas which is returned to the `interstellar medium of the system'. Thus, we can write
(see, e.g., \citealp{tinsley73})

\begin{equation}
\frac{d^{2} M_{\rm ej}}{dVdt} = \int_{m(t)}^{\rm 140{\rm M}_\odot}{(m-m_{\rm r})\Phi(m)\Psi(t-\tau_{m})dm},\label{mej1}
\end{equation}

\noindent where the limit $m(t)$ corresponds to the stellar mass whose lifetime is equal to $t$. In the integrand, $m_{\rm r}$ is
the mass of the remnant, which depends on the progenitor mass (see \citealp{tinsley73} for details), and the star formation rate is
taken at the retarded time $(t-\tau_{\rm m})$, where $\tau_{\rm m}$ is the lifetime of a star of mass $m$. 

For all stars formed in the haloes, it is used the metallicity-independent fit of \citet{s6,c4}

\begin{equation}
 \log_{10}(\tau_{\rm m})=10.0-3.6\,\log_{10}\left(\frac{M}{\rm M_{\odot}}\right) +\left[ \log_{10}
\left( \frac{M}{\rm M_{\odot}}\right) \right]^{2},
\end{equation}

\noindent where $\tau_{\rm m}$ is the stellar lifetime given in years.

In equation (\ref{mej1}), the term $\Phi(m)$ represents the initial mass function (IMF) which gives the distribution function
of stellar masses. Thus,

\begin{equation}
\Phi(m) = A m^{-(1+x)}\label{imf1},
\end{equation}

\noindent where $x$ is the slope of the IMF, and $A$ is a normalization factor determined by

\begin{equation}
\int_{0.1{\rm M}_\odot}^{140{\rm M}_\odot}m\Phi(m)dm = 1.\label{norimf}
\end{equation}

Numerical integration of (\ref{rhogas}) produces the function $\rho_{\rm g}(t)$ at each time $t$ (or redshift $z$).
Once obtained $\rho_{\rm g}(t)$, we return to Eq. (\ref{sclaw}) in order to obtain the cosmic star formation rate (CSFR).
Just replacing $\Psi(t)$ by $\dot\rho_{\star}(t)$, we have

\begin{equation}
\dot\rho_{\star}=k\rho_{\rm g}\label{csfr}.
\end{equation}

It is worth stressing that although we did not take into account the stellar feedback processes on the derivation of the
CSFR (see, e.g., \citealp{crs10} for this issue), our models, as we will discuss below, have good agreement with observational data 
taken from \citet{h2,h3} at lower redshifts ($z < 5$). Furthermore, the CSFR obtained in the present work has good agreement with that one derived by 
\citet{sh2003} from hydrodynamic simulations. Regardless, we should comment on this limitation of the model in its present form. In particular, stellar feedback
processes can modify the time-scale ($\tau_{\rm s}$) for star formation, for example, through radiation, winds, and supernova events from massive stars. As a  main
result, the star formation efficiency, $\varepsilon_{\star}$, embedded in the normalization of the CSFR at $z=0$ (see \citealp{pereira10} for this issue) and
$\tau_{\rm s}$ would be functions of time. In principle, $\tau_{\rm s}$ and $\varepsilon_{\star}$ not constants can modify the shape of the CSFR at higher redshifts.
On the other hand, due to the good agreement with observations at $z<2$, where data are less scattered, the values used in this work should represent reasonable
mean values of these quantities over the whole interval $[0,z_{\rm ini}]$. Certainly, the inclusion of stellar feedback processes would be an interesting refinement
to introduce in future works.

\subsection{the dark-energy models and the input parameters}

The last point we have to consider for the characterization of the CSFR is the dark-energy component of the Universe through its
equation-of-state parameter $w(a)$. In the present work, we consider two cases: cosmological constant where $w(a) = -1$ and
the Chaplygin gas.

In particular, the Chaplygin gas is characterized by a fluid with an equation-of-state $p=-A /\rho^{\alpha}$. This dark energy fluid has been tested
against observational data as, for example, SNIa (e.g, \citealp{fabris05}), cosmic microwave background (e.g., \citealp{pia10}), and power spectrum
(e.g., \citealp{fabris10}) and it configures in a strong alternative candidate to the cosmological constant.

Its equation-of-state parameter is given by

\begin{equation}
 w(a) = -\frac{A}{A+B\,a^{-3(\alpha+1)}}.\label{chap}
\end{equation}

In Eq. (\ref{chap}), the constants $A$ and $B$ are (see, e.g., \citealp{pace10} for details)

\begin{equation}
 A = -w_{0}(\Omega_{\rm d}\rho_{\rm c})^{1+\alpha}\;\;\;\; {\rm and} \;\;\;\; B = (1+w_{0})(\Omega_{\rm d}\rho_{\rm c})^{1+\alpha}.
\end{equation}

The present value of the equation-of-state parameter is related to $A$ and $B$ by

\begin{equation}
 w_{0} = - \frac{A}{A+B}.
\end{equation}

The cosmological parameters we have used in this work are: $\Omega_{\rm d}= 0.762$, $\Omega_{\rm m} = 0.238$,
$\Omega_{\rm b} = 0.042$, and Hubble constant $H_{0}=100\,h\,{\rm km}\,{\rm s}^{-1}\,{\rm Mpc}^{-1}$ with $h=0.734$.
For the variance of the overdensity field smoothed on a scale of size $8\,h^{-1}\,{\rm Mpc}$ we consider $\sigma_{8} =0.8$.
The parameters associated to the Chaplygin gas are $\alpha = 1.0$ (classical Chaplygin gas) and $\alpha = 0.2$ (generalized
version). In both cases we consider $w_{0} = -0.8$.

In the column 1 of Table 1 is shown the name of the models. In column 2 we present the slope of the initial mass
function ($x$ in equation \ref{imf1}), the time-scale for star formation is presented in column 3, the redshift
($z_{\rm p}$) where the CSFR peaks is presented in column 4, and finally in column 5 we have the kind of dark fluid.

All models presented in Table 1 have good agreement with observational data. In particular, it was performed $\chi^{2}$ analysis over
these models, obtaining the reduced chi-square defined as $\chi_{\rm r} = \chi^{2}/{\rm dof}$ (where ``dof" means ``degrees of freedom").
All of these models satisfy $\chi_{\rm r} < 1$. In Fig. 1 we present the CSFR derived from equation (\ref{csfr}) for three models of
Table 1. The observational points are taken from \citet{h2,h3}.

\begin{table}
{\center
\caption{The input parameters used to obtain the Cosmic Star Formation Rates. All CSFRs have good agreement with observational data.
The redshift $z_{\rm ini}$ associated with the beginning of star formation is $20$. In the fifth column, $\Lambda$ represents the
cosmological constant and $\alpha$ represents the Chaplygin gas.}
 \label{tab1}
 \begin{tabular}{@{}lcccc}
  \hline
 CSFR & $x\ (\rm IMF)$ & $\tau_{\rm s}\ {\rm Gyr}$ & $z_{\rm p}$ & dark fluid \\
 \hline
 A1  &  1.35  &  2.0  &  3.54  &  $\Lambda$       \\
 A2  &  1.35  &  3.0  &  2.94  &  $\Lambda$       \\
 A3  &  0.35  &  1.0  &  3.29  &  $\Lambda$       \\
 A4  &  1.35  &  2.0  &  2.75  &  $\alpha = 0.2$  \\
 A5  &  1.35  &  3.0  &  2.21  &  $\alpha = 0.2$  \\
 A6  &  0.35  &  1.0  &  2.52  &  $\alpha = 0.2$  \\
 A7  &  1.35  &  2.0  &  2.42  &  $\alpha = 1.0$  \\
 A8  &  1.35  &  3.0  &  1.91  &  $\alpha = 1.0$  \\
 A9  &  0.35  &  1.0  &  2.21  &  $\alpha = 1.0$  \\
 \hline
\end{tabular}

\medskip
}
\end{table}

\begin{figure}
\includegraphics[width=90mm]{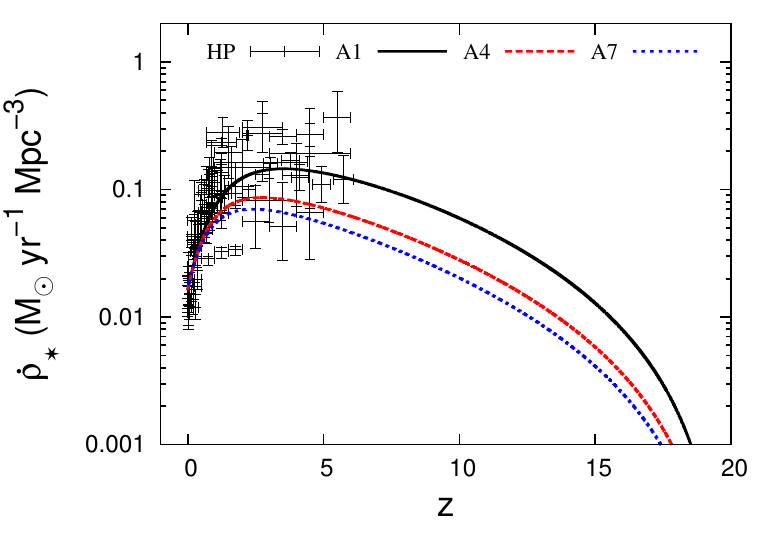}
\caption{The CSFR derived in this work compared to the observational points (HP) taken from \citet{h2,h3}.
The models are described in Table 1.}
\end{figure}

We can see from the results in Table 1 and Fig. 1 that the cosmological constant produces higher amplitudes than the Chaplygin gas.
That means, the process of baryonic matter infall from the haloes is more efficienty, for the same set of parameters, if the dark
energy fluid is the cosmological constant. Another characteristic which can be seen from Fig. 1 and Table 1 is that the Chaplygin
gas decreases the redshift where the CSFR peaks when compared to the cosmological constant. 

\section[]{The coalescence rates}

We assume that the coalescence rates track the CSFR but with a delay $t_{\rm d}$ between the formation of the binary
system and the final merger (see \citealp{regimbau09}). Thus, we can write

\begin{equation}
\dot\rho_{\rm c}^{0}(z) = \dot\rho_{\rm c}^{0}(0)\times \frac{\dot\rho_{\star c}(z)}{\dot\rho_{\star c}(0)},\label{coal1}
\end{equation}

\noindent where $\dot\rho_{\rm c}^{0}(z)$ is the rate at which binary systems are observed to merger at redshift $z$, and $\dot\rho_{\rm c}^{0}(0)$
is the same rate in our local Universe.

The connection between the past CSFR and the rate of binary merger is given by $\dot\rho_{\star c}(z)$ through
the relation

\begin{equation}
\dot\rho_{\star c}(z)=\int_{\tau_{0}}^{t(z)}\frac{\dot\rho_{\star}(z_{\rm f})}{(1+z_{\rm f})}P(t_{\rm d})dt_{\rm d},\label{coal2}
\end{equation}

\noindent where $\dot\rho_{\star}(z_{\rm f})$ is the CSFR obtained from Eq. (\ref{csfr}), $P(t_{\rm d})$ is the
probability per unit of time of merging after the formation of the progenitor, including both the evolutionary time for
the formation of the compact binary and the time for the compact binary to coalesce, and the $(1+z_{\rm f})$ term in
the denominator considers the time dilatation due to the cosmic expansion.

The time delay $t_{\rm d}$ makes the connection between the redshift $z$ at which a compact binary system mergers, and
the redshift $z_{\rm f}$ at which its progenitor was formed. As discussed by Regimbau \citet{regimbau09}, it can be calculate by

\begin{equation}
t_{\rm d} = \frac{1}{H_{0}}\int_{z}^{z_{\rm f}} \frac{dz'}{(1+z')E(z')}.\label{delay}
\end{equation}

The probability $P(t_{\rm d})$ is described of the form

\begin{equation}
P(t_{\rm d})\propto \frac{1}{t_{\rm d}},
\end{equation}

As mentioned by Regimbau \& Hughes (2009), this form accounts for the wide range of merger times observed in binary pulsars.
This form was also used by de Freitas Pacheco (1997), \citet{regimbau}, and \citet{pach2} in their works. Thus, we define $P(t_{\rm d})$ as

\begin{equation}
P(t_{\rm d})=\frac{B}{t_{\rm d}},
\end{equation}

\noindent where $B$ is a normalization constant, and the probability function $P(t_{\rm d})$ is normalized in the range of $\tau_{0}-15\,{\rm Gyr}$
for some minimal delay $\tau_{0}$. Therefore,

\begin{equation}
\int_{\tau_{0}}^{15{\rm Gyr}} \frac{B}{t_{\rm d}}dt_{\rm d} = 1.
\end{equation}

Specifically, we consider that $\tau_{0} = 20\, {\rm Myr}$ for NS-NS systems, $\tau_{0} = 10\, {\rm Myr}$ for NS-BH systems, and
$\tau_{0} = 100\, {\rm Myr}$ for BH-BH systems (\citealp{bul04}).

With these assumptions, the merger rate per unit of redshift can be written as

\begin{equation}
\frac{dR_{\rm c}^{0}}{dz} = \dot\rho_{\rm c}^{0}(z)\frac{dV}{dz},\label{rate}
\end{equation}

\noindent where $dV$ is the comoving volume element which is given by

\begin{equation}
dV = 4\pi r(z)^{2} \frac{c}{H_{0}}\frac{dz}{E(z)}.\label{volume}
\end{equation}

In Eq. (\ref{volume}), $r(z)$ is the proper distance, whose expression is

\begin{equation}
r(z) = \frac{c}{H_{0}}\int_{0}^{z}\frac{dz}{E(z)}.\label{distprop}
\end{equation}

Note that the expansion function $E(z)$ is dependent on the kind of dark-energy fluid as shown by Eq. (\ref{expan}). Then, using
the formalism described in this Section, it can be determined the cosmic coalescence rates for NS-NS binaries,
NS-BH, and BH-BH systems up to redshift $z\sim 20$. Figure 2 presents the cosmic coalescence rates, normalized to the local value
$\dot\rho_{\rm c}^{0}(0)$, for NS-NS binaries. On the other hand, Figures 3 and 4 respectively show the cosmic coalescence rates
for NS-BH and BH-BH binaries.

\begin{figure}
\includegraphics[width=90mm]{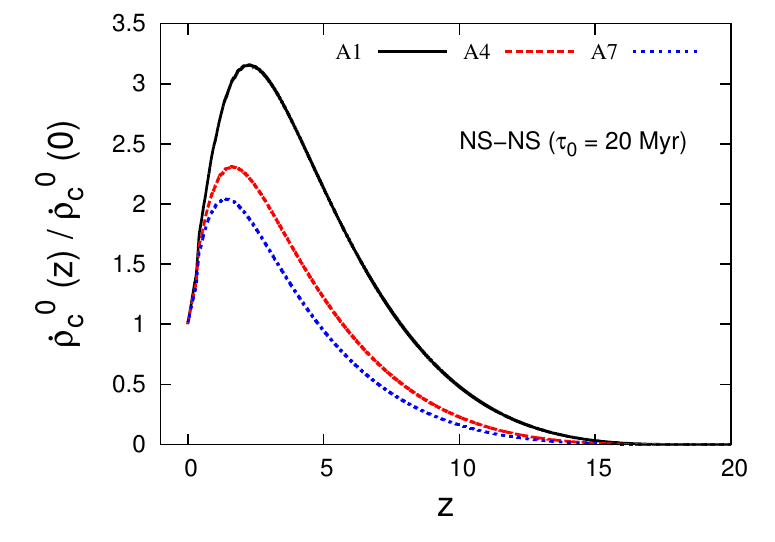}
\caption{The cosmic coalescence rates for NS-NS binaries. The models consider $\tau_{0} = 20\,{\rm Myr}$.
These models are obtained from the CSFR presented in Fig. 1 (see also Table 1 for the main parameters).}
\end{figure}

\begin{figure}
\includegraphics[width=90mm]{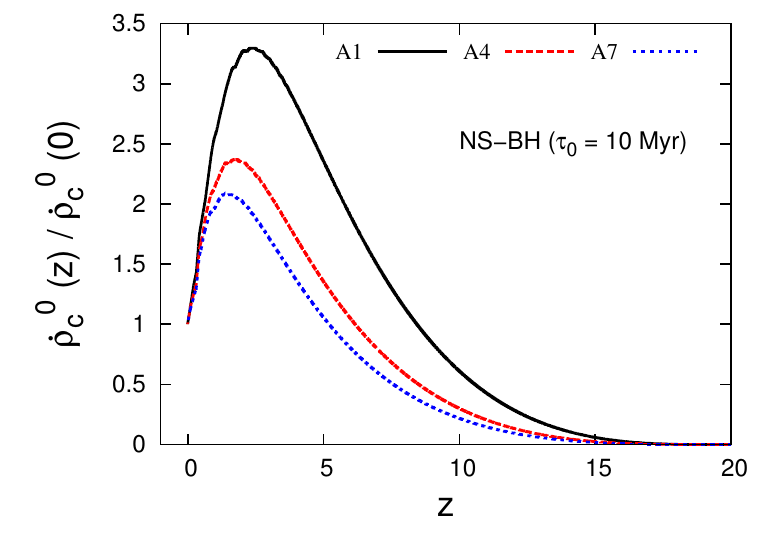}
\caption{The cosmic coalescence rates for NS-BH binaries. The models consider $\tau_{0} = 10\,{\rm Myr}$.
These models are obtained from the CSFR presented in Fig. 1 (see also Table 1 for the main parameters).}
\end{figure}

\begin{figure}
\includegraphics[width=90mm]{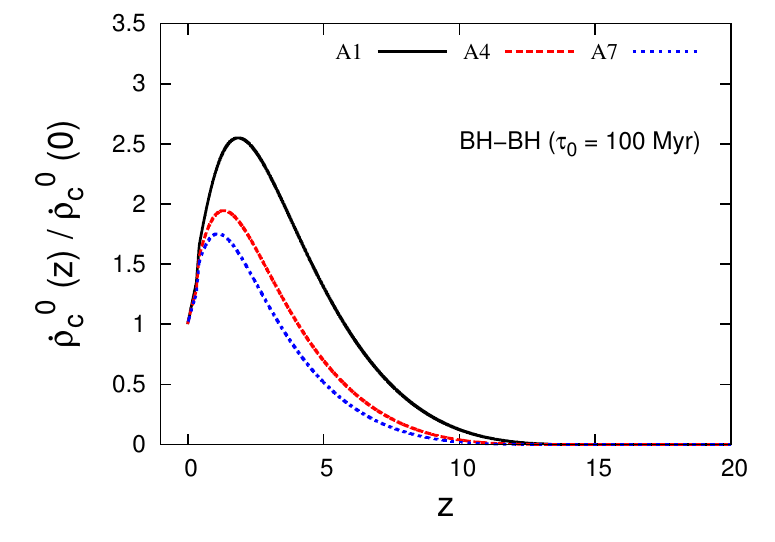}
\caption{The cosmic coalescence rates for BH-BH binaries. The models consider $\tau_{0} = 100\,{\rm Myr}$.
These models are obtained from the CSFR presented in Fig. 1 (see also Table 1 for the main parameters).}
\end{figure}

As expected, due to the behavior of the CSFR, the amplitudes of the coalescence rates produced by the
Chaplygin gas cosmology are lower than those produced by the cosmological-constant cosmology. In particular,
for the cosmological constant (model A1) $\dot\rho_{\rm c}^{0}(z)/\dot\rho_{\rm c}^{0}(0)$ reaches a maximum amplitude at
redshift $z=2.27$ for NS-NS binaries, at $z=2.45$ for NS-BH and at $z=1.86$ for BH-BH systems. In the case
generalized Chaplygin gas ($\alpha =0.2$ $-$ model A4) we note that $\dot\rho_{\rm c}^{0}(z)/\dot\rho_{\rm c}^{0}(0)$ peaks
at $z=1.63$ for NS-NS, at $z=1.81$ for NS-BH and at $z=1.31$ for BH-BH. The last case, classical Chaplygin gas ($\alpha = 1.0$),
reaches a maximum amplitude at $z=1.34$ for both NS-NS and NS-BH systems. On the other hand, for BH-BH binaries the maximum
value of the coalescence rate is reached at $z=1.13$. Thus, the position of the peak of the coalescence rate is dictated by
the value of $\tau_{0}$ and also by the kind of dark-energy fluid.

As we will see in the next Section, the different behaviors for the coalescence rates produced by different dark-energy cosmologies
will produce different values for the signal-to-noise ratios of advanced LIGO and Einstein Telescope. 

\section[]{Gravitational wave background}

\subsection[]{Compact binary systems}

The spectrum of a stochastic background of GWs is characterized by the closure energy density per logarithmic frequency spam,
which is given by (see, e.g., \citealp{a1,a2})

\begin{equation}
\Omega_{\rm GW} = \frac{1}{\rho_{\rm c}} \frac{{\rm d}\rho_{\rm GW}}{{\rm d}\log \, \nu_{\rm obs}}, \label{omegagw}
\end{equation}

\noindent where $\rho_{\rm GW}$ is the gravitational energy density, and $\nu_{\rm obs}$ is the frequency in the observer frame.

The above equation can be written as (see, e.g., \citealp{f2})

\begin{equation}
\Omega_{\rm GW} = \frac{1}{c^{3}\rho_{\rm c}}\nu_{\rm obs}F_{{\nu}_{\rm obs}},
\end{equation}

\noindent where $F_{{\nu}_{\rm obs}}$ is the gravitational wave flux (given in ${\rm erg}\, {\rm cm}^{-2}\, {\rm Hz}^{-1}\, {\rm s}^{-1}$)
at the observer frequency $\nu_{\rm obs}$ integrated over all cosmological sources. Therefore,

\begin{equation}
F_{{\nu}_{\rm obs}} = \int_{0}^{z_{\rm ini}} f_{{\nu}_{\rm obs}}\, dR_{\rm c}^{0}(z).\label{flux}
\end{equation}

See that $dR_{\rm c}^{0}(z)/dz$ is the merger rate per unit of redshift (Eq. \ref{rate}). In order to solve Eq. (\ref{flux}), it is
needed to determine the gravitational wave fluence ($f_{{\nu}_{\rm obs}}$), in the observer frame, produced by a given compact binary
coalescence. Following \citet{regimbau,regimbau11,zhu11,mar3,rosado11,wu12,regimbau12}, $f_{{\nu}_{\rm obs}}$ can be written as

\begin{equation}
f_{{\nu}_{\rm obs}} = \frac{1}{4\pi d_{\rm L}^{2}}\frac{dE_{\rm GW}}{d\nu}(1+z)^{2},\label{fluence}
\end{equation}

\noindent where $d_{\rm L} = r(z)(1+z)$ is the luminosity distance, $r(z)$ is the proper distance (see Eq. \ref{distprop}),
$dE_{\rm GW}/{d\nu}$ is the spectral energy, and $\nu=\nu_{\rm obs}(1+z)$ is the frequency in the source frame.

In the quadrupolar approximation, the spectral energy emitted by a compact binary system, with masses $m_{1}$ and $m_{2}$, which
inspirals in a circular orbit is given by \citep{peters63}

\begin{equation}
\frac{dE_{\rm GW}}{d\nu} = K\, \nu^{-1/3},
\end{equation}

\noindent where

\begin{equation}
K = \frac{(G\pi)^{2/3}}{3}\frac{m_{1}m_{2}}{(m_{1}+m_{2})^{1/3}}.
\end{equation}

It will be considered that the GW background has the value of the maximum frequency limited by the `last stable orbit' (LSO).
Then, following \citet{sathya} 

\begin{equation}
\nu_{\rm max} = \nu_{\rm LSO} = 1.5\left(\frac{M}{2.8\,{\rm M}_{\odot}}\right)^{-1} \, {\rm kHz},
\end{equation}

\noindent where $M$ is the total mass of the system ($M=m_{1}+m_{2}$). 

In the present study, we consider $m_{1}=m_{2}=1.4\,{\rm M}_{\odot}$ for NS-NS binaries, while for NS-BH are used
$m_{1}=1.4\,{\rm M}_{\odot}$ and $m_{2}=7.0\,{\rm M}_{\odot}$. Concerning to BH-BH systems we have used $m_{1}=m_{2}=7.0\,{\rm M}_{\odot}$.
With these considerations, the maximum frequency is $\nu_{\rm LSO}=1.5\,{\rm kHz}$ ($500\,{\rm Hz}$) for NS-NS (NS-BH). For BH-BH
binaries we have $\nu_{\rm LSO}=300\,{\rm Hz}$.

There is one last point to consider before calculating the spectrum of the stochastic background of GWs. This point is related to the value
of the local merger rate per unit volume $\dot\rho_{\rm c}^{0}(0)$. As discussed by \citet{regimbau09}, the local merger is usually extrapoled
by multiplying the rate in the Milk Way with the density of equivalent galaxies.

Current estimates give $\dot\rho_{\rm c}^{0}(0)=(0.01-10)\times {\rm Myr^{-1}Mpc^{-3}}$ for NS-NS, and
$\dot\rho_{\rm c}^{0}(0)=(0.001-1)\times {\rm Myr^{-1}Mpc^{-3}}$ for NS-BH (see \citealp{regimbau09} and the references therein).
In the present work, it is considered $\dot\rho_{\rm c}^{0}(0)= 1.0\, {\rm Myr^{-1}Mpc^{-3}}$ for NS-NS,
$\dot\rho_{\rm c}^{0}(0)= 0.1\, {\rm Myr^{-1}Mpc^{-3}}$ for NS-BH, and
$\dot\rho_{\rm c}^{0}(0)= 0.01\, {\rm Myr^{-1}Mpc^{-3}}$ for BH-BH.

Thus, using the formalism above, we can obtain the characterization of the stochastic background of GWs formed by
the coalescence of compact binary systems. In particular, the Figures 5, 6, and 7 present the spectra of the
gravitational energy density parameter $\Omega_{\rm GW}$ versus the observed frequency $\nu_{\rm obs}$. 

\begin{figure}
\includegraphics[width=90mm]{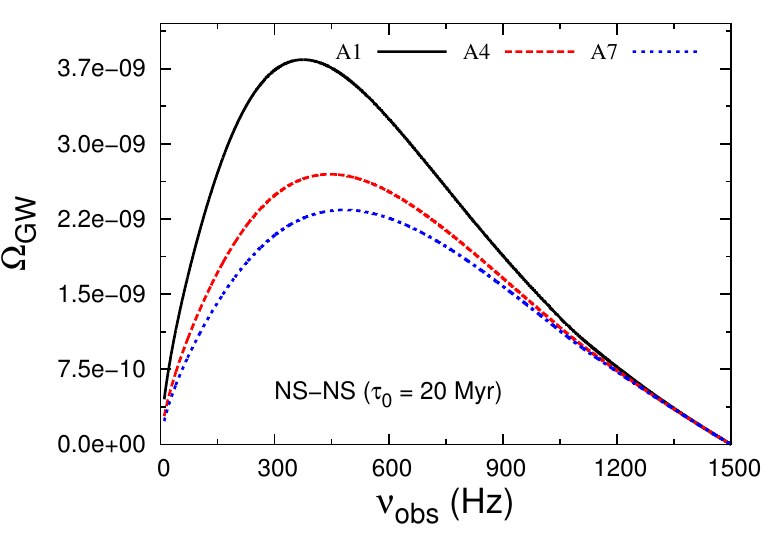}
\caption{Spectrum of the gravitational energy density parameter $\Omega_{\rm GW}$. The results are shown for double neutron stars (NS-NS).}
\end{figure}

\begin{figure}
\includegraphics[width=90mm]{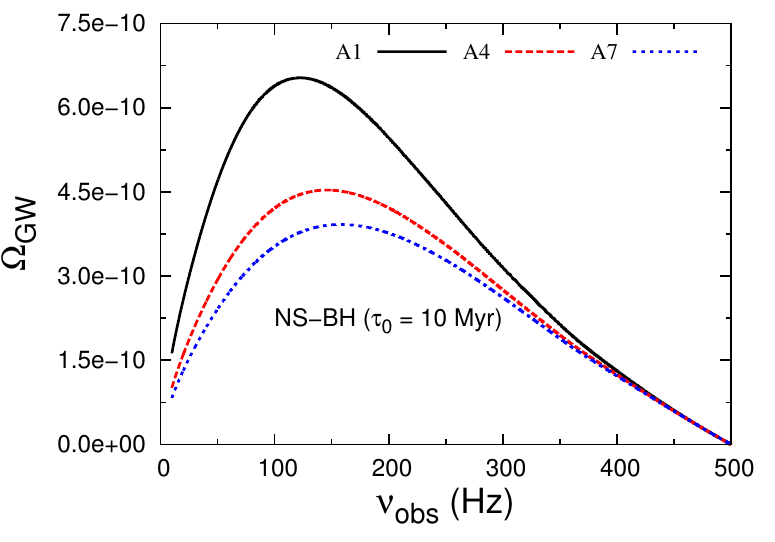}
\caption{Spectrum of the gravitational energy density parameter $\Omega_{\rm GW}$. The results are shown for the coalescence of
neutron star-black hole (NS-BH) systems.}
\end{figure}

\begin{figure}
\includegraphics[width=90mm]{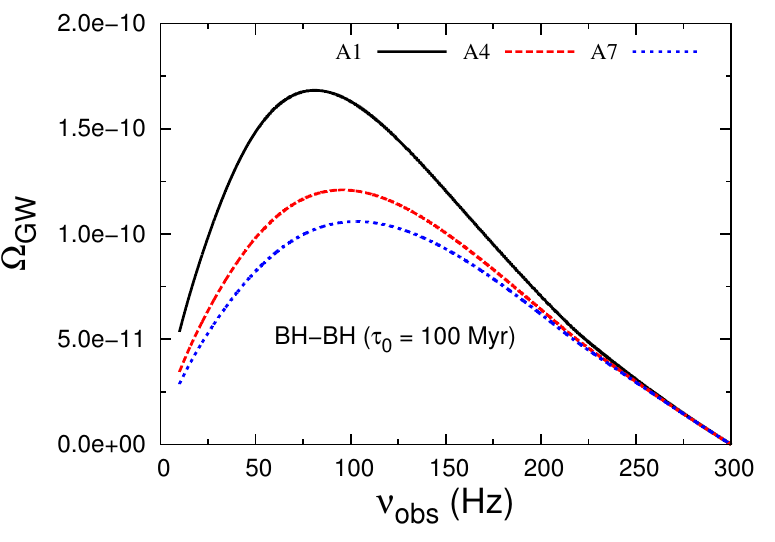}
\caption{Spectrum of the gravitational energy density parameter $\Omega_{\rm GW}$. The results are shown for the coalescence of
black hole-black hole (BH-BH) systems.}
\end{figure}

The density parameter increases as $\nu_{\rm obs}^{2/3}$ at low frequencies and reaches a maximum amplitude
$\sim 3.8\times 10^{-9}$ around $375\, {\rm Hz}$ for NS-NS systems if is used the CSFR-A1 (cosmological constant).
It is worth stressing that calculations performed by \citet{regimbau}, using Monte Carlo methods for obtaining the coalescence rates,
produced similar results.

In particular, these authors obtained maximum amplitude of about $1.1\times 10^{-9}$ around $670\,{\rm Hz}$ for NS-NS binaries
(considering their fiducial CSFR). However, observing the distribution of coalescences as a function of the redshift derived by
\citet{regimbau} we note that it peaks at $z\sim 1.5$. On the other hand, our coalescence rates peak at
$z\sim 2.9-3.5$ (models A1 to A3 which correspond to the cosmological constant as dark-fluid). In this way, the maximum value
of $\Omega_{\rm GW}$ is shifted to lower frequency than that obtained by \citet{regimbau}.

In order to assess the detectability of a gravitational wave signal, one must evaluate the signal-to-noise ratio (S/N), which
for a pair of interferometers is given by (see, for example, \citealp{crs,f5,a1,d4,d5,regimbau})

\begin{equation}
{\rm (S/N)^{2}} = \left[\left({\frac{9H_{0}^{4}}{50\pi^{4}}}\right) T\int_{0}^{\infty}d\nu\frac{\gamma^{2}(\nu)
\Omega_{\rm GW}^{2}(\nu)}{\nu^{6}S_{\rm h}^{(1)}(\nu)S_{\rm h}^{(2)}(\nu)}\right],\label{signal}
\end{equation}

\noindent where $S_{\rm h}^{(i)}$ is the spectral noise density, $T$ is the integration time, and $\gamma(\nu)$ is the
overlap reduction function, which depends on the relative positions and orientations of the two interferometers. For
the $\gamma(\nu)$ function we refer the reader to \citet{f5} who was the first to calculate a closed form for the LIGO
observatories.

The noise power spectral density of the advanced LIGO can be found at LIGO website. However, we used here the analytical fit given by \citet{mis10}. Its expression is

\begin{eqnarray}
S_{\rm h}(\nu)&=&S_0\bigg[10^{16-4 (\nu-7.9)^2}+2.4\times10^{-62}\,x^{-50}+0.08\,x^{-4.69} \nonumber\\
 &+&  123.35\,\left(\frac{1-0.23\,x^2+0.0764\,x^4}{1+0.17\,x^2}\right)\bigg]\;\; {\rm if} \;\; \nu \geq \nu_{\rm s}
\,,\nonumber\\
&=&\infty \;\; {\rm if} \;\; \nu< \nu_{\rm s},
\label{adv_ligo}
\end{eqnarray}

\noindent where $x=\nu/\nu_{0}$, $\nu$ stands for the frequency, $\nu_{0}=215\,{\rm Hz}$, $S_{0} = 10^{-49}
\,{\rm Hz}^{-1}$, and $\nu_{\rm s}$ is a low-frequency cut-off that can be varied, and below which $S_{\rm h}(\nu)$
can be considered infinite for all practical purposes (here, we choose $\nu_{\rm s}=10\,{\rm Hz}$).

On the other hand, a possibility for a third generation ground-based gravitational wave detector is the Einstein
Telescope (ET). The basic design of this inteferometer is still under discussion so there exist some possible
sensitivity curves\footnote{See also http://www.et-gw.eu/etsensitivities. In particular, the ET-C and ET-D
sensitivity curves correspond to a xylophone configuration consisting of a pair of detectors. The first one
operating at low-frequency $(1-100\,{\rm Hz})$ and the second detector operating at high-frequency
$(100-\sim 10\,{\rm kHz})$.} (\citealp{hild08,hild10,punt10a,punt10b,sathya12}). 

Here, we use the ET-B sensitivity curve of \citet{hild08} with an analytical fit taken from \citet{mis10}.

\begin{eqnarray}
S_{\rm h}(\nu) & = & S_0 \left[a_1 x^{b_1}+a_2 x^{b_2}
+a_3 x^{b_3}+a_4 x^{b_4}\right]^{2}\; {\rm if} \; \nu \geq \nu_{\rm s}\, \nonumber\\
&=& \infty \;\;\;\; {\rm if} \;\; \nu < \nu_{\rm s},\label{et_b}
\end{eqnarray}

\noindent where $x=\nu/\nu_{0}$, $\nu$ stands for the frequency, $\nu_{0}=100\,{\rm Hz}$,
$S_{0} = 10^{-50}\,{\rm Hz}^{-1}$, and $\nu_{\rm s}$ is a low-frequency cut-off that can be varied, and below which
$S_{\rm h}(\nu)$ can be considered infinite for all practical purposes (here, we choose $\nu_{\rm s}=10\,{\rm Hz}$).
The coefficients in Eq. (\ref{et_b}) have the values

\begin{eqnarray}
a_1 &=&2.39\times10^{-27},\qquad\, b_1=-15.64,\nonumber\\
a_2 &=&0.349, \,~~~~\qquad\qquad b_2=-2.145,\nonumber\\
a_3 &=&1.76, ~~~\qquad\qquad\quad b_3=-0.12,\nonumber\\
a_4 &=&0.409, ~~\qquad\qquad\quad b_4=1.10.
\label{constants_sqrtpsd}
\end{eqnarray}

We consider that the ET has a triangular configuration (\citealp{hild10}) with an overlap reduction function 
given by \citet{regimbau11}. In Tables 2, 3, and 4 we summarize the main characteristics of the models. We show
the values of the maximum amplitude ($\Omega_{\rm GW_{max}}$) of the stochastic background, the frequency ($\nu_{\rm p}$)
where $\Omega_{\rm GW}$ peaks, and the signal-to-noise ratios for advanced LIGO and ET interferometers. Note that for all kind of compact binaries we have ${\rm (S/N)} \lesssim 1$ for a pair of advanced LIGOs. On the other hand,
for ET in triangular configuration should be possible, in principle, to obtain high values for (S/N).

\begin{table}
{\center
\caption{The main characteristics of the models NS-NS and their respectives signal-to-noise ratios (S/N) for a pair of
`advanced LIGOs' and Einstein Telescope in triangular configuration. The integration time is $T = 1\, {\rm yr}$. We also
show the redshift $z_{\rm DC}$ at which the duty cycle becomes equal to 1 (transition between the popcorn and
the continuous stochastic regime).}
\label{tab2}
 \begin{tabular}{@{}lccccc}
  \hline
 NS-NS & $\Omega_{\rm GWmax}$ & $\nu_{\rm p}\, ({\rm Hz})$ & ${\rm(S/N)}$ &  ${\rm(S/N)}$ & $z_{\rm DC}$  \\
       &                      &                            &  Adv. LIGO    &   ET    &  $D=1$  \\ 
 \hline
 A1  &  $3.84\times 10^{-9}$  &  375  &  1.46  & 335 & 0.55  \\
 A2  &  $3.11\times 10^{-9}$  &  415  &  1.11  & 256 & 0.57  \\
 A3  &  $3.57\times 10^{-9}$  &  390  &  1.32  & 305 & 0.56  \\
 A4  &  $2.70\times 10^{-9}$  &  446  &  0.92  & 211 & 0.61  \\
 A5  &  $2.27\times 10^{-9}$  &  485  &  0.71  & 168  & 0.64  \\
 A6  &  $2.53\times 10^{-9}$  &  461  &  0.84  & 194  & 0.62  \\
 A7  &  $2.34\times 10^{-9}$  &  482  &  0.76  & 174  & 0.64  \\
 A8  &  $2.00\times 10^{-9}$  &  521  &  0.61  & 142  & 0.67  \\
 A9  &  $2.21\times 10^{-9}$  &  498  &  0.70  & 161  & 0.65  \\
 \hline
\end{tabular}

\medskip
}
\end{table}

\begin{table}
{\center
\caption{The main characteristics of the models NS-BH and their respectives signal-to-noise ratios (S/N) for a pair of
`advanced LIGOs' and Einstein Telescope in triangular configuration. The integration time is $T = 1\, {\rm yr}$. We also
show the redshift $z_{\rm DC}$ at which the duty cycle becomes equal to 0.1 (transition between the shot noise and
the popcorn regime).}
 \label{tab3}
 \begin{tabular}{@{}lccccc}
  \hline
 NS-BH & $\Omega_{\rm GWmax}$ & $\nu_{\rm p}\, ({\rm Hz})$ & ${\rm(S/N)}$ &  ${\rm(S/N)}$ & $z_{\rm DC}$  \\
       &                      &                            &  Adv. LIGO     &   ET    &  $D=0.1$ \\ 
 \hline
 A1  &  $6.53\times 10^{-10}$  &  123  &  0.52  & 116  & 0.93  \\
 A2  &  $5.25\times 10^{-10}$  &  136  &  0.40  & 88  & 0.99  \\
 A3  &  $6.06\times 10^{-10}$  &  128  &  0.47  & 105  & 0.94  \\
 A4  &  $4.53\times 10^{-10}$  &  146  &  0.33  & 73  & 1.15  \\
 A5  &  $3.79\times 10^{-10}$  &  159  &  0.26  & 58  & 1.28  \\
 A6  &  $4.25\times 10^{-10}$  &  151  &  0.30  & 67  & 1.19  \\
 A7  &  $3.92\times 10^{-10}$  &  158  &  0.27  & 61  & 1.30  \\
 A8  &  $3.35\times 10^{-10}$  &  172  &  0.22  & 49  & 1.51  \\
 A9  &  $3.69\times 10^{-10}$  &  164  &  0.25  & 56  & 1.37  \\
 \hline
\end{tabular}

\medskip
}
\end{table}

\begin{table}
{\center
\caption{The main characteristics of the models BH-BH and their respectives signal-to-noise ratios (S/N) for a pair of
`advanced LIGOs' and Einstein Telescope in triangular configuration. The integration time is $T = 1\, {\rm yr}$.}
 \label{tab4}
 \begin{tabular}{@{}lcccc}
  \hline
 BH-BH & $\Omega_{\rm GWmax}$ & $\nu_{\rm p}\, ({\rm Hz})$ & ${\rm(S/N)}$ &  ${\rm(S/N)}$ \\
       &                      &                            & Adv. LIGO    &   ET    \\ 
 \hline
 A1  &  $1.68\times 10^{-10}$  &  81  &  0.17  & 36    \\
 A2  &  $1.40\times 10^{-10}$  &  89  &  0.14  & 29    \\
 A3  &  $1.58\times 10^{-10}$  &  84  &  0.16  & 33    \\
 A4  &  $1.21\times 10^{-10}$  &  96  &  0.11  & 24    \\
 A5  &  $1.04\times 10^{-10}$  &  103  &  0.09  & 20    \\
 A6  &  $1.14\times 10^{-10}$  &  99  &  0.10  & 22    \\
 A7  &  $1.06\times 10^{-10}$  &  103  &  0.09  & 20    \\
 A8  &  $9.21\times 10^{-11}$  &  110  &  0.08  & 17    \\
 A9  &  $1.00\times 10^{-10}$  &  106  &  0.09  & 19    \\
 \hline
\end{tabular}

\medskip
}
\end{table}

Concerning the nature of the GW background, it is determined by the duty cycle which is defined as the ratio, in the
observer frame, of the typical duration of a single burst $\bar \tau$ to the average time interval between sucessive
events (see \citealp{regimbau,regimbau09})

\begin{equation}
D(z) = \int_{0}^{z} {\bar \tau}\frac{dR_{\rm c}^{0}}{dz'}dz'\label{duty},
\end{equation}

\noindent where

\begin{equation}
{\bar \tau} = \frac{5c^{5}}{256\pi^{8/3}G^{5/3}}\left[\left(1+z'\right) m_{\rm c}\right]^{-5/3}f_{\rm L}^{-8/3}\label{duty2},
\end{equation}

\noindent with $f_{\rm L}$ being the lower frequency bound of the detector, and $m_{\rm c}$ represents the chirp mass
which is given by

\begin{equation}
m_{\rm c} = \frac{(m_{1}m_{2})^{3/5}}{(m_{1}+m_{2})^{1/5}}.\label{chirp}
\end{equation}

Figure 8 presents $D(z)$ for the NS-NS binaries while Fig. 9 shows the duty cycle for NS-BH systems. In Figure 10 we have
the duty cycle for BH-BH systems. In these plots we have considered $f_{\rm L}=10\,{\rm Hz}$.

Concerning the duty cycle, there are three different regimes for this parameter (see, e.g., \citealp{rosado11,wu12,regimbau12}). The first case appears when $D(z) < 0.1$. In this case, we have the so-called `shot noise regime' consisting of a sequency of widely spaced events. That means the sources can be resolved individually.

The second case appears when $0.1 < D(z) < 1$. We have the `popcorn noise regime' That means the time interval between
two sucessive events could be closer to the duration of a single event. In reality, near to $D(z)=1$ the events may
overlap, making it difficult to identify individual events.

The third case appears when $D(z)\geq 1$. In this case, we have a `continuous background'. The signals overlap to
produce a continuous stochastic background.

In Table 2 we include the value of the redshift $z_{\rm DC}$ at which the background becomes continuous
($D(z)>1$). In Table 3 we present the redshift of transition between the shot noise and the popcorn regime
($D(z)>0.1$). In particular, for NS-NS binaries a continuous background is established for sources situated at
cosmological distances $z\sim 0.5-0.6$ ($z\sim 0.6-0.7$) for the cosmological constant (Chaplygin gas) cosmology.

On the other hand, for the NS-BH binaries the signals change from shot noise to popcorn regime at $z\sim 0.9-1.0$
($z\sim 1.0-1.5$ for Chaplygin gas). For the BH-BH systems the signals are always within the shot noise regime.
However, note that reducing the values of the local merger $\dot\rho_{\rm c}^{0}(0)$ in relation to those values
used in the present work, it would reduce the values of the (S/N) besides changing the regimes (or values of $D(z)$)
of the gravitational wave backgrounds. In particular, see that (S/N) and $D(z)$ are proportional to
$\dot\rho_{\rm c}^{0}(0)$.

\begin{figure}
\includegraphics[width=90mm]{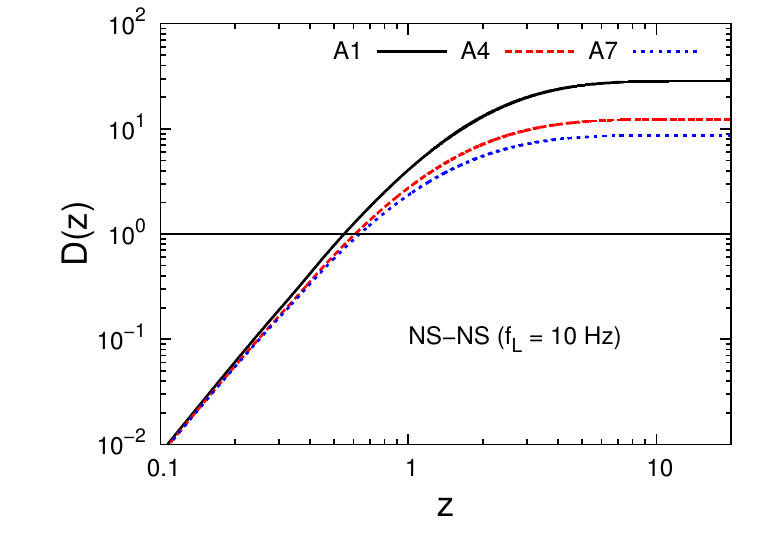}
\caption{The duty cycle as a function of the redshift for NS-NS binaries. We consider $f_{\rm L}=10\,{\rm Hz}$.
The horizontal line at $D(z)=1$ represents the transition between the popcorn regime and the continuous stochastic
background.}
\end{figure}

\begin{figure}
\includegraphics[width=90mm]{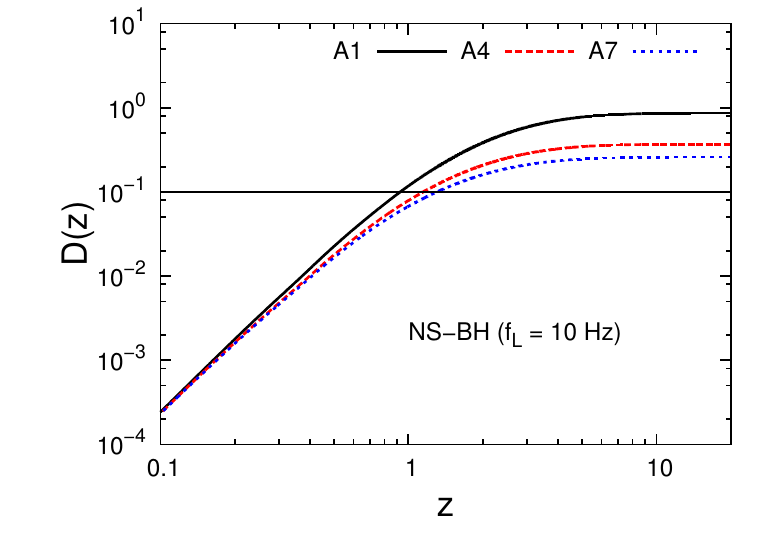}
\caption{The duty cycle as a function of the redshift for NS-BH binaries. We consider $f_{\rm L}=10\,{\rm Hz}$.
The horizontal line at $D(z)=0.1$ represents the transition between the shot noise and the popcorn regime.}
\end{figure}

\begin{figure}
\includegraphics[width=90mm]{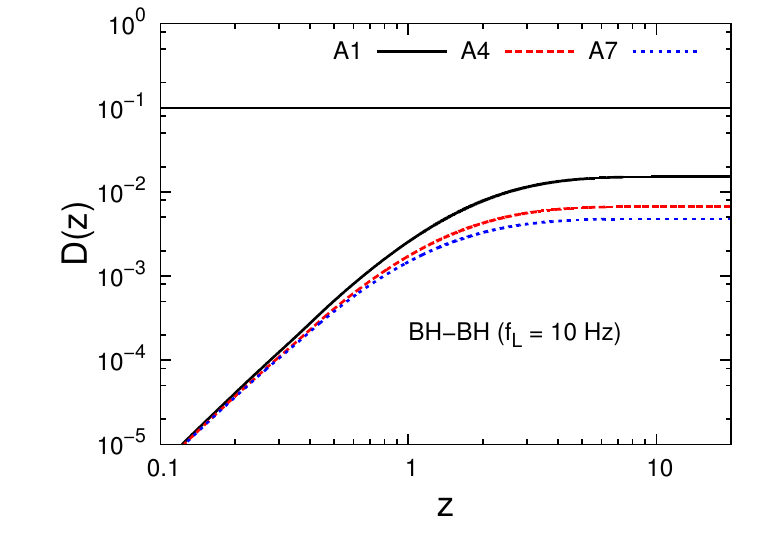}
\caption{The duty cycle as a function of the redshift for BH-BH binaries. We consider $f_{\rm L}=10\,{\rm Hz}$.
These compact binaries produce a signal of the kind shot noise.}
\end{figure}

\subsection[]{Collapse of stars to form  black holes}

In order to determine the background of GWs generated by stars which collapse to black holes, we re-write
Eq. (\ref{flux}) as

\begin{equation}
F_{{\nu}_{\rm obs}} = \int_{0}^{z_{\rm ini}} f_{{\nu}_{\rm obs}}\, dR_{\rm BH}(z),\label{flux2}
\end{equation}

\noindent where now we have

\begin{equation}
\frac{dR_{\rm BH}}{dz} = \dot\rho_{\star}(z)\Phi(m)dm\frac{dV}{dz}\label{dR_BH},
\end{equation}

\noindent and for $f_{{\nu}_{\rm obs}}$ we have

\begin{equation}
f_{{\nu}_{\rm obs}} = \frac{\pi c^{3}}{2G}h_{\rm BH}^{2}.\label{flux3}
\end{equation}

The dimensionless amplitude $h_{\rm BH}$ is given by (\citealp{t1})

\begin{equation}
h_{\rm BH} \simeq 7.4 \times 10^{-20} \varepsilon_{\rm GW}^{1/2}\left(\frac{m_{\rm r}}
{\rm M_{\sun}}\right)\left(\frac{d_{\rm L}}{1\rm{Mpc}}\right)^{-1},\label{h_BH}
\end{equation}

\noindent where $\varepsilon_{\rm GW}$ is the efficiency of generation of GWs, and $m_{\rm r}$ is the mass of the
black hole formed.

It is worth mentioning that Eq. (\ref{h_BH}) refers to the black hole `ringing', which has to do with the de-excitation
of the black hole quasi-normal modes.

The collapse of a star to black hole produces a signal with frequency $\nu_{\rm obs}$ given by

\begin{equation}
\nu_{\rm obs} \simeq 1.3 \times 10^{4} {\rm Hz}\left(\frac{\rm{M}_{\sun}}{m_{\rm r}}\right)(1+z)^{-1}.
\end{equation}

We will consider that black holes are formed from stars with $40\ {\rm M}_{\odot} \leq m \leq 140\ {\rm M}_{\odot}$.
The lower limit is consistent with recent results derived from the X-ray pulsar CXO J164710.2$-$455216 which show that
the progenitor to this pulsar had an initial mass $\sim 40\,{\rm M}_{\odot}$ (\citealp{muno06}). On the other hand,
the mass of the black hole remnant is taken to be the mass of the helium core before collapse (see \citealt{h1}).
Thus,

\begin{equation}
m_{\rm r}=m_{\rm He}=\frac{13}{24}(m-20\ {\rm M}_{\sun}).\label{mr}
\end{equation} 

With these considerations, we can obtain the spectrum of GWs produced by cosmological black holes. The Fig. 11 shows
the spectrum of the gravitational energy density parameter $\Omega_{\rm GW}$ as a function of the observed frequency
$\nu_{\rm obs}$ for the models with the highest (S/N) of Table 5. These curves consider
$\varepsilon_{\rm GW}=10^{-4}$ (\citealp{lof06}).

We can see that the spectra peak at $\Omega_{\rm GW} \sim 3\times 10^{-9}-10^{-7}$ dependent on both CSFR parameters
and dark-energy component. Note that only two models have transition from shot noise to popcorn regime
(which corresponds to $D=0.1$). All the other models correspond to shot noise signals.  

\begin{figure}
\includegraphics[width=90mm]{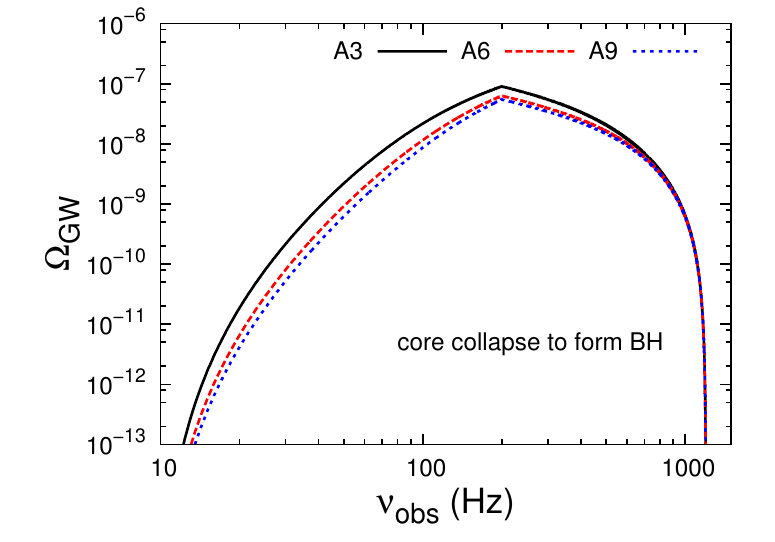}
\caption{Spectrum of the gravitational energy density parameter $\Omega_{\rm GW}$. We consider an efficiency of
generation of gravitational waves $\varepsilon_{\rm GW} = 10^{-4}$.}
\end{figure}

\begin{table}
{\center
\caption{The main characteristics of the models `core-collapse to form black holes'. The efficiency of generation
of GWs is $\varepsilon_{\rm GW}=10^{-4}$. The signal-to-noise ratios (S/N) for a pair of `advanced LIGO' are
determined for an integration time $T = 1\, {\rm yr}$. For ET we consider triangular configuration. We also
show the redshift $z_{\rm DC}$ at which the duty cycle becomes equal to 0.1 (transition from shot noise to
popcorn regime).}
 \label{tab5}
 \begin{tabular}{@{}lccccc}
  \hline
 BH & $\Omega_{\rm GWmax}$ & $\nu_{\rm p}\, ({\rm Hz})$ & ${\rm(S/N)}$ &  ${\rm(S/N)}$  & $z_{\rm DC}$ \\
       &                      &                            & Adv. LIGO    &   ET  &  $ D = 0.1$ \\
 \hline
 A1  &  $6.80\times 10^{-9}$  &  200  &  0.14  &  152  &  $-$ \\
 A2  &  $5.00\times 10^{-9}$  &  200  &  0.08  &  96   &  $-$ \\
 A3  &  $9.02\times 10^{-8}$  &  200  &  1.76  &  1900 &  $3.38$ \\
 A4  &  $4.70\times 10^{-9}$  &  200  &  0.07  &  86   &  $-$ \\
 A5  &  $3.63\times 10^{-9}$  &  200  &  0.04  &  58   &  $-$ \\
 A6  &  $6.29\times 10^{-8}$  &  200  &  0.85  &  1100 &  $6.92$ \\
 A7  &  $4.06\times 10^{-9}$  &  200  &  0.05  &  68   &  $-$ \\
 A8  &  $3.20\times 10^{-9}$  &  200  &  0.03  &  48   &  $-$ \\
 A9  &  $5.46\times 10^{-8}$  &  200  &  0.62  &  870  &  $-$ \\
 \hline
\end{tabular}

\medskip
}
\end{table}

\begin{figure}
\includegraphics[width=90mm]{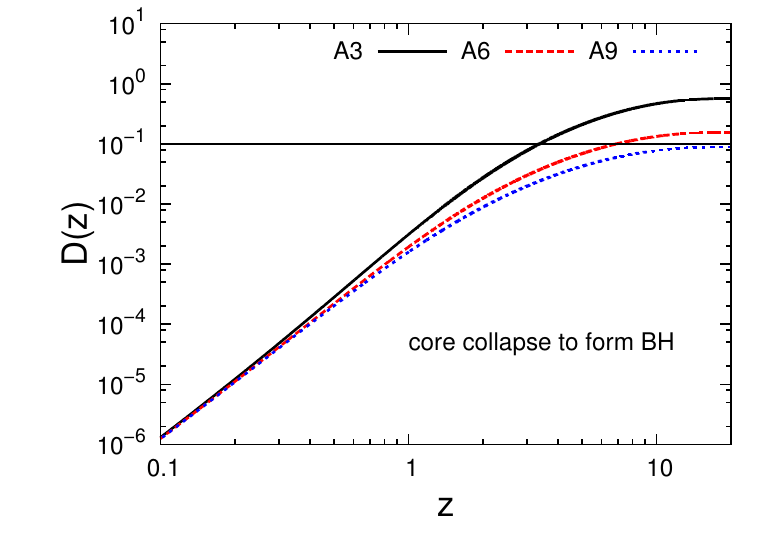}
\caption{The duty cycle as a function of the redshift for the cosmological Population of black holes. The horizontal
line at $D(z)=0.1$ represents the transition from shot noise to popcorn regime.}
\end{figure}

The Fig. 12 presents the duty cycle generated by the collapse of stars to form black holes. Note that in this case
we calculate the duty cycle as

\begin{equation}
D(z) = \int_{0}^{z} {\bar \tau}(1+z')\frac{dR_{\rm BH}}{dz'}dz'\label{duty3},
\end{equation}

\noindent with $\bar\tau = 1\,{\rm ms}$ (\citealp{f2}).

We can observe from Table 5 that the frequency where the spectra peak neither depend on the dark-energy cosmology
nor on the CSFR parameters. Another characteristic of this kind of source is the very high (S/N) produced for ET.

It is worth stressing that \cite{zhu10} has recently estimated the upper limit on the energy density, of a stochastic gravitational wave background, produced by the core-collapse supernovae leaving black holes as remnants. The authors showed that considering Gaussian source spectra would be possible to detect GW signals with $\varepsilon_{\rm GW} \sim 10^{-5}$ $(10^{-7})$ for Adv. Ligo (ET). Another work centerded on the GW backgrounds produced by core-collapse supernovae of Population III and Population II stars was developed by \cite{mar1}. In particular, the authors studied the cosmic transition of Population III to Population II using waveforms derived from recent 2D numerical simulations. The GW efficiencies used by these authors were $\varepsilon_{\rm GW} \sim 10^{-7}$ for Population II progenitors and $\varepsilon_{\rm GW} \sim 10^{-5}$ for Population III progenitors (with initial masses ranging between $100\, {\rm M}_\odot$ to $500\, {\rm M}_\odot$). Here, we have adopted the GW spectrum of \cite{t1}, with efficiency $\varepsilon_{\rm GW} \sim 10^{-4}$ because the core-collapse energy spectrum will affect the results for both cases, $\Lambda$-CDM and Chaplygin gas, exactly the same way.

\section[]{Cosmological spectrum produced by all sources}

It is worth stressing that a stochastic background of gravitational waves is expected to arise from a superposition
of a large number of gravitational wave sources of astrophysical and cosmological origin. In particular, the
Wilkinson Microwave Anisotropy Probe (WMAP) results suggest an early epoch for the reionization of the Universe
(see, e.g., \citealp{j2011}). In this way, a pre-galactic Population should be formed at high redshifts to account for
these results. The cosmic star formation history is determined by the interplay between the incorporation of baryons into
collapsed objects and return of baryons into diffuse state (e.g., gaseous clouds).

Thus, the formation of different objects as, for example, NS-NS binaries, NS-BH binaries, BH-BH systems, core-collapse
to form black holes, among others, are directly related to the CSFR. On the other hand, different dark-energy scenarios
could give different signatures on the background through the expansion function $E(z)$ (Eq. \ref{expan}). As the 
gravitational wave background could trace the behavior of the Universe up to redshift $\sim 20$, then it
should be possible to infer if there is a temporal dependence of the dark-energy equation of state. That is, if
$\dot w(a) \neq 0$.

In this way, the detection and characterization of a stochastic background of GWs could be used as a tool for the
study of the Universe at high redshifts. In particular, the gravitational wave signals produced at different
cosmological distances by the sources discussed above could overlap at a given frequency $\nu_{\rm obs}$ to produce a
stochastic background over a large range in frequency.

In Figure 13 we show the collective contribution of the three compact binary sources investigated here. In Table 6
we summarize the main characteristics of these models. Although the (S/N) of the collective spectra are dominated by
the NS-NS binaries, we note that the NS-BH and BH-BH binaries pull the peak of the collective spectra for lower frequencies than those observed if we only consider the NS-NS systems.

\begin{figure}
\includegraphics[width=90mm]{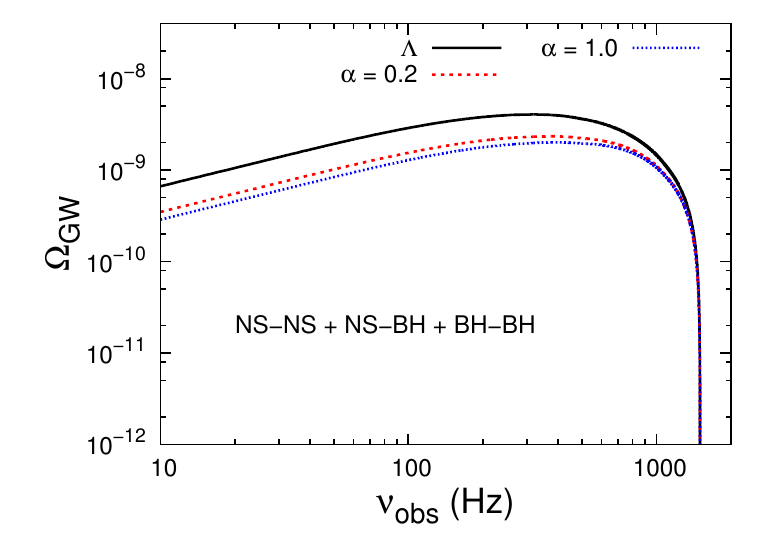}
\caption{Collective spectra of the three compact binary sources studied in this work. The curves represent the models A1, A4, and A7.}
\end{figure}

\begin{table}
{\center
\caption{The main characteristics of the collective contribution of the three compact binary sources, namely,
NS-NS binaries, NS-BH systems and BH-BH binaries. The signal-to-noise ratios (S/N) for a pair of `advanced LIGO'
are determined for an integration time $T = 1\, {\rm yr}$. We also present the (S/N) for ET in triangular configuration.}
 \label{tab6}
 \begin{tabular}{@{}lcccc}
  \hline
Model & $\Omega_{\rm GWmax}$ & $\nu_{\rm p}\, ({\rm Hz})$ & ${\rm(S/N)}$ &  ${\rm(S/N)}$ \\
       &                      &                            & Adv. LIGO    &   ET  \\
 \hline
 A1  &  $4.06\times 10^{-9}$  &  321  &  2.15  &  486 \\
 A2  &  $3.27\times 10^{-9}$  &  346  &  1.64  &  373 \\
 A3  &  $3.77\times 10^{-9}$  &  331  &  1.95  &  442 \\
 A4  &  $2.81\times 10^{-9}$  &  374  &  1.35  &  308 \\
 A5  &  $2.34\times 10^{-9}$  &  398  &  1.08  &  246 \\
 A6  &  $2.63\times 10^{-9}$  &  385  &  1.24  &  283 \\
 A7  &  $2.42\times 10^{-9}$  &  396  &  1.12  &  254 \\
 A8  &  $2.04\times 10^{-9}$  &  413  &  0.91  &  207 \\
 A9  &  $2.27\times 10^{-9}$  &  404  &  1.03  &  235 \\
 \hline
\end{tabular}

\medskip
}
\end{table}

In Figure 14 we include the core-collapse supernovae together the compact binary systems in the calculation
of the gravitational wave spectra. We can see that same for a low efficiency of generation of gravitational waves
($\varepsilon_{\rm GW}=10^{-4}$) the collective spectra show a clear signature of this kind of source when compared
to the spectra only derived with compact binaries.

\begin{figure}
\includegraphics[width=90mm]{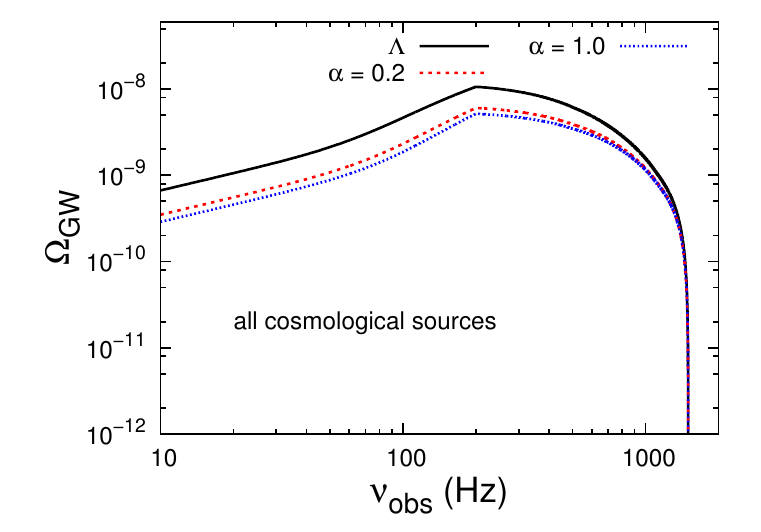}
\caption{Spectrum of all sources studied in this work. Concerning to the core-collapse to form black holes, we use
$\varepsilon_{\rm GW}=10^{-4}$. The curves represent the models A1, A4, and A7.}
\end{figure}

In Table 7 we present the main caractheristics of the collective spectra with all sources studied in this work.

\begin{table}
{\center
\caption{The main characteristics of the collective contribution of all sources studied in this work. The signal-to-noise
ratios (S/N) for a pair of `advanced LIGO' are determined for an integration time $T = 1\, {\rm yr}$. We also present
the (S/N) for ET in triangular configuration.}
 \label{tab7}
 \begin{tabular}{@{}lcccc}
  \hline
Model & $\Omega_{\rm GWmax}$ & $\nu_{\rm p}\, ({\rm Hz})$ & ${\rm(S/N)}$ &  ${\rm(S/N)}$ \\
       &                      &                            & Adv. LIGO    &   ET  \\
 \hline
 A1  &  $1.06\times 10^{-8}$  &  200  &  2.23  &  590 \\
 A2  &  $8.00\times 10^{-9}$  &  200  &  1.69  &  433 \\
 A3  &  $9.36\times 10^{-8}$  &  200  &  3.26  &  2182 \\
 A4  &  $7.23\times 10^{-9}$  &  200  &  1.39  &  362 \\
 A5  &  $5.69\times 10^{-9}$  &  200  &  1.10  &  280 \\
 A6  &  $6.52\times 10^{-8}$  &  200  &  1.81  &  1262 \\
 A7  &  $6.19\times 10^{-9}$  &  200  &  1.14  &  295 \\
 A8  &  $4.96\times 10^{-9}$  &  200  &  0.92  &  233 \\
 A9  &  $5.65\times 10^{-8}$  &  200  &  1.42  &  1005 \\
 \hline
\end{tabular}

\medskip
}
\end{table}

We note that core-collapse supernovae have an important contribution for the shape of the collective spectrum for
frequencies $60-\sim 300\,{\rm Hz}$. In particular, the frequency where the collective spectra peak is completely
dominated by the core-collapse supernovae. See that model A3 could produce $({\rm S/N})\sim 3$ ($\sim 2200$) for
advanced LIGO (ET) in the cosmological-constant cosmology. On the other hand, model A6 could produce
$({\rm S/N})\sim 2$ ($\sim 1300$) for advanced LIGO (ET) in the Chaplygin gas cosmology. In principle, detecting
stochastic backgrounds of gravitational waves with high (S/N) could be possible to infer, for example, the behavior
of the CSFR at high-redshifts as well as if there is a temporal dependence of the dark-energy equation
of state.

\section{Compact binaries and core-collapse parameters: Influence on the results}

In the previous Sections, we have analyzed the main characteristics of the stochastic backgrounds produced by compact binary systems,
core-collapse to form black holes, and the composite signal of these cosmological objects. We verify that higher signal-to-noise ratios
can be produced if we consider Einstein Telescope in triangular configuration. Although these cosmological sources are connected by the CSFR,
we know that there are uncertainties in the minimum coalescence time-scales of NS-NS, NS-BH, and BH-BH binaries. On the other hand, the local
coalescence rates of these systems can vary up to three orders of magnitude. In addition, the minimum mass able to form a black hole may vary from
$\sim 25{\rm M}_{\odot}$ to $\sim 40{\rm M}_{\odot}$. Thus, in this Section, we present an analysis of these uncertainties and their influence on the
stochastic backgrounds discussed here. We also discuss if there is a clear difference between the $\Lambda$CDM and the Chaplygin gas which would
permit constrain both the constant feature (or not) of the dark-energy equation of state and the CSFR derived for $\Lambda$CDM and Chaplygin gas. In
particular, we have analyzed:

\begin{figure}
\includegraphics[width=90mm]{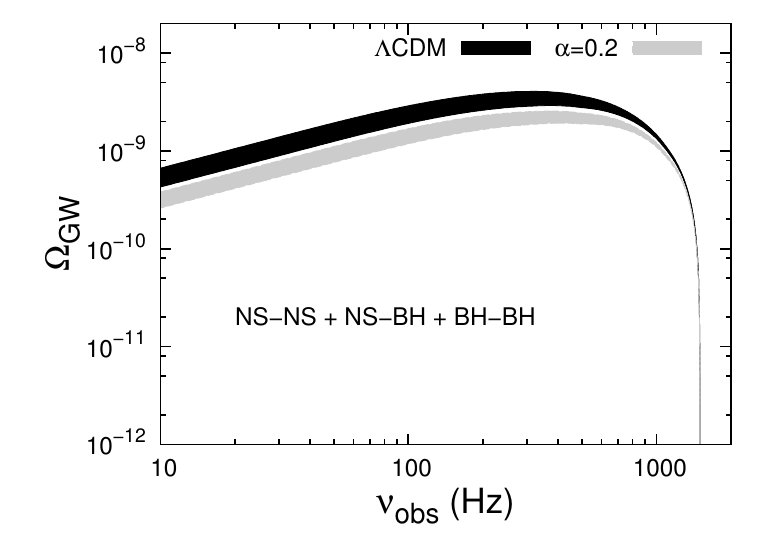}
\caption{Collective spectra of the three compact binaries taking into account the uncertainties in the parameters. The black area describes all the possible
GW signals for the $\Lambda$CDM case. The gray area represents the GW backgrounds for Chaplygin gas with $\alpha=0.2$.}
\end{figure}

a) The local coalescence rate: This parameter acts like an offset and it does not modify the shapes of the spectra. Note, however,
that ${\rm (S/N)}\propto \dot\rho_{\rm c}^{0}(0)$ and so our values for the signal-to-noise ratios can vary from 0.1 to 10 of those listed in Tables $2-4$.
Thus, this parameter can only modify the values of the (S/N) as ${\rm (S/N)} = \dot\rho_{\rm c}^{0}(0)/\dot\rho_{\rm c}^{0}(0)_{u}\times
{\rm (S/N)}_{u}$ where the subscript $u$ means the values used and derived in this work. Note that same in the worst case ($\dot\rho_{\rm c}^{0}(0)
=0.1\times \dot\rho_{\rm c}^{0}(0)_{u})$, it would be possible to have ${\rm (S/N)}> 10$ (ET) for the composite signals of these three binary sources.
Concerning to the duty cycle, observe that $D(z)$ is also proportional to $\dot\rho_{\rm c}^{0}(0)$. Thus, the redshifts of transition from popcorn to continuous regimes
(and from shot noise to popcorn regimes) can change according to the values of $\dot\rho_{\rm c}^{0}(0)$;

b) The minimum stellar mass to form a black hole: This parameter basically changes the maximum frequency of the background formed by core-collapse.
In the case $m_{\rm min} = 25{\rm M}_{\odot}$, we obtain $\nu_{\rm max}=4.8\,{\rm kHz}$ while for $m_{\rm min} = 40{\rm M}_{\odot}$ we have $\nu_{\rm max}=
1.2\,{\rm kHz}$. In terms of (S/N), if we change $m_{\rm min}$ from $40{\rm M}_{\odot}$ to $25{\rm M}_{\odot}$ the signal-to-noise ratios increase in
$5\%$ in relation to those values present in Table 5. Looking for the results in Table 7, collective contribution of all sources, (S/N) increases by
$2\%$ ($50\%$) for Adv. LIGO (ET). The frequency where $\Omega_{\rm GW}$ peaks is weakly dependent on this parameter in both cases $\Lambda$CDM and Chaplygin gas;

c) Efficiency of generation of gravitational waves ($\varepsilon_{\rm GW}$): Note that $\Omega_{\rm GW} \propto \varepsilon_{\rm GW}$. Thus, same with an efficiency
of generation of GWs $\sim 10^{-5}-10^{-6}$ could be possible to have $({\rm S/N})> 10$ for ET in triangular configuration (see Table 5);

\begin{figure}
\includegraphics[width=90mm]{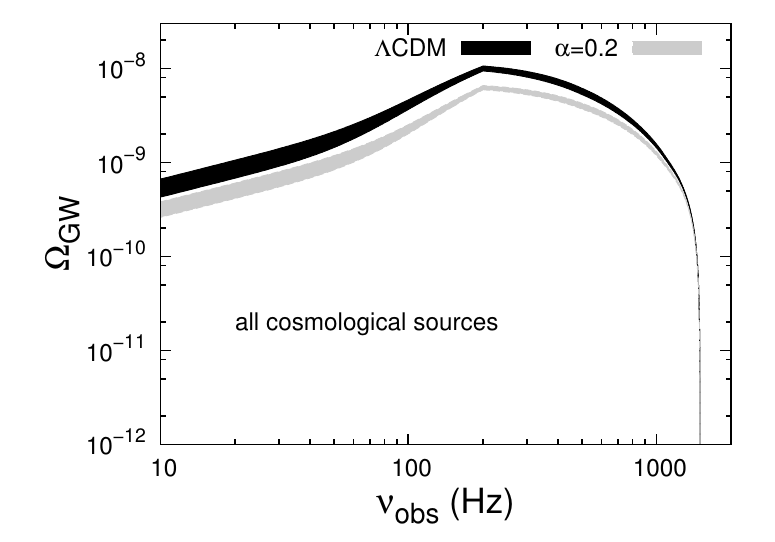}
\caption{Collective spectra of all sources studied in this work and taking into account the uncertainties in the parameters. The black area describes all
the possible GW signals for the $\Lambda$CDM case. The gray area represents the GW backgrounds for Chaplygin gas with $\alpha=0.2$.}
\end{figure}

d) Coalescence time-scale of NS-NS: We change this parameter from $20\,{\rm Myr}$ to $100\,{\rm Myr}$. As a consequence, the coalescence rate peaks at
$z\sim 1.9\, (1.30)$ instead of $z\sim 2.27\, (1.63)$ for the $\Lambda$CDM (Chaplygin gas with $\alpha = 0.2$) while the (S/N) of the Table 2
typically decreases by $15\%$. Looking for the collective contribution of all sources in Table 7 we note that (S/N) decreases by $8\%$ ($3\%$) for Adv. LIGO (ET).
There is just a slight modification of the frequency where $\Omega_{\rm GW}$ peaks.

e) Coalescence time-scale of NS-BH: We change this parameter from $10\,{\rm Myr}$ to $50\,{\rm Myr}$. As a consequence, the coalescence rate peaks at
$z\sim 2.1\, (1.50)$ instead of $z\sim 2.45\, (1.81)$ for the $\Lambda$CDM (Chaplygin gas with $\alpha = 0.2$) while the (S/N) of the Table 3
typically decreases by $10\%$. Looking for the collective contribution of all sources in Table 7 we note that (S/N) decreases by $1\%$ for both Adv. LIGO
and ET.

f) Coalescence time-scale of BH-BH: We change this parameter from $100\,{\rm Myr}$ to $500\,{\rm Myr}$. As a consequence, the coalescence rate peaks at
$z\sim 1.25\, (0.83)$ instead of $z\sim 1.86\, (1.31)$ for the $\Lambda$CDM (Chaplygin gas with $\alpha = 0.2$) while the (S/N) of the Table 4
typically decreases by $20\%$. Looking for the collective contribution of all sources in Table 7 we note that (S/N) decreases by $0.8\%$ for both Adv. LIGO
and ET.

A question could arise about the uncertainties described above: Is it possible to have a clear separation of the two backgrounds
($\Lambda$CDM and Chaplygin gas cosmologies) or the uncertainties listed above produce a superposition of these backgrounds? A
second question could also arise: Can different dark energy scenarios produce distinct signatures on the CSFR? In order to answer
these questions, we present in Figures 15 and 16 the gravitational wave backgrounds with the uncertainties
discussed above and for the models A1 and A3 of Table 1. In these Figures, we just keep fixed two parameters: $\varepsilon_{\rm GW}=10^{-4}$ and
$m_{\rm min}=40\,{\rm M}_{\odot}$. Below $\nu_{\rm obs}\sim 1\, {\rm kHz}$ there is no superposition between the GW signals in the case $\Lambda$CDM and
Chaplygin gas ($\alpha = 0.2$), same with all the uncertainties in the parameters. However, note that the case Chaplygin gas with $\alpha =1$ can not be separated
from $\alpha=0.2$. There is a superposition between these two Chaplygin models if we take into account all the uncertainties discussed above.

\begin{figure}
\includegraphics[width=90mm]{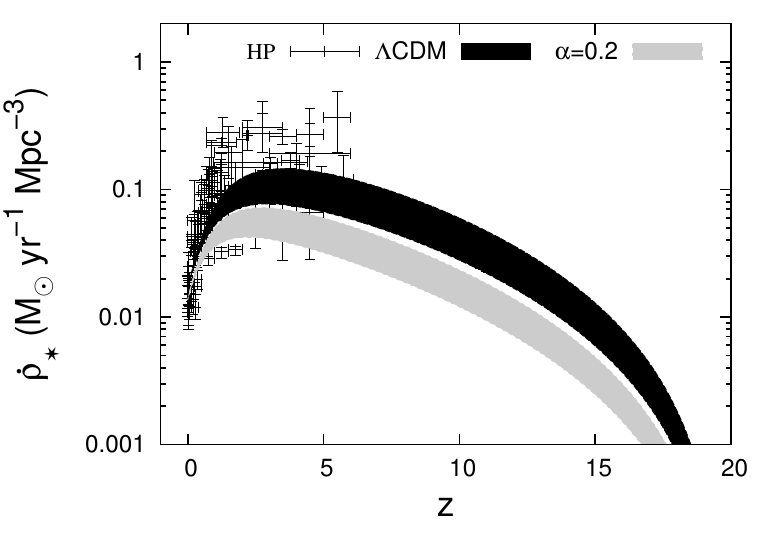}
\caption{Possible CSFRs taking into account all the viable models studied in this work. The black are represents the family of CSFRs for the
$\Lambda$CDM cosmology (models A1 to A3) while the gray area shows the family of CSFRs for the Chaplygin gas (models A4 to A6) as dark-energy component
of the Universe. Note that there is no overlap at $z> 2$ between these two dark-fluids same with all the uncertainties in the parameters.}
\end{figure}

The second point is related to the CSFR. Looking for Figure 17, we see that there is no overlap between the case $\Lambda$CDM and Chaplygin gas at $z>2$
same with all the uncertainties in the parameters. The areas defined by $\Lambda$CDM and Chaplygin gas cosmology do not overlap in the redshift range $[2-20]$.
In principle, with observational data less scattered in the range $z\sim 2-5$, it would be possible to have a better indication of the
dark-energy equation of state from the observed CSFR. Otherwise, being detected a stochastic background of GWs with high (S/N), as in the case of ET,
we could work with the inverse problem reconstructing the CSFR from the observed background. In this way, ET could contribute with a better comprehension
of how star formation is regulated at high redshifts.

\section{Final remarks}

In this work, we have first studied the main characteristics of the gravitational wave signals produced by coalescences
of NS-NS, NS-BH, and BH-BH binaries up to redshift $z\sim 20$. The coalescence rates are obtained from the hierarchical
formation scenario recently studied by \citet{pereira10}.

In this formalism, the `cosmic star formation rate' (CSFR)
is derived in a self-consistent way, considering the baryon accretion rate as an infall term which supplies the gaseous
reservoir in the haloes. However, here we modify their model in order to incorporate different dark-energy fluids.
In particular, we show results considering two different dark-energy components of the Universe: the cosmological constant
and the Chaplygin gas.

For NS-NS systems, the shape of the spectrum of the gravitational energy density parameter ($\Omega_{\rm GW}$) has good
agreement with the case studied by \citet{regimbau} who used numerical simulations based on Monte Carlo methods.
In particular, we have obtained signal-to-noise ratios $({\rm S/N})\sim 1.5$ for NS-NS, $({\rm S/N})\sim 0.50$
for NS-BH, and $({\rm S/N})\sim 0.20$ for BH-BH binaries in the cosmological-constant cosmology and considering
the correlation of two Adv. LIGO detectors. If we consider ET in triangular configuration, the (S/N) are at least a factor
$\sim 200$ greater than those obtained for Adv. LIGO. The signals produced in the case Chaplygin gas are always lower than those produced
by the cosmological-constant cosmology.

We have also analyzed the nature of the GW background produced by those compact binaries. For our fiducial parameters, we verify
that a continuous background, corresponding to a duty cycle $\gtrsim 1$, is produced by sources situated at cosmological distances
far from $z\sim 0.5-0.6$ for NS-NS binaries in the cosmological-constant cosmology. For the Chaplygin gas cosmology,
duty cycle $\gtrsim 1$ is obtained for sources far from $z\sim 0.6-0.7$.

Considering NS-BH binaries, the nature of the background becomes popcorn for sources far from $z\sim 0.9-1.0$
($z\sim 1.0-1.5$) if the dark component of the Universe is the cosmological constant (Chaplygin gas). On the other hand,
BH-BH binaries are always within the shot noise regime.

We also verify the characteristics of the background produced by a cosmological Population of stellar
core-collapse to form black holes. In this case, signal-to-noise ratios within the range from
$\sim 0.05$ to $\sim 2$ could be generated depending upon the CSFR/dark-energy component and the efficiency of
generation of GWs used. We verify that this Population does not behaves as a continuous background. However, for
sources situated at cosmological distances $z\sim 3$ ($z\sim 7$) we verify a transition from shot noise to
popcorn regime if the time-scale for star formation ($\tau_{\rm s}$) is $\sim 1\,{\rm Gyr}$ and if the
dark-energy component is the cosmological constant (Chaplygin gas). 

A stochastic background of GWs, from cosmological origin, is expected to arise from a superposition of different
gravitational wave sources at different redshifts. In particular, the formation of different objects as, for example,
NS-NS binaries, NS-BH binaries, BH-BH systems, core-collapse supernovae to form black holes, among others, are related
to the CSFR. In this way, we determine the shape of a pre-galactic background of GWs considering the collective effect
of these different objects. We obtain that a stochastic background of GWs could be generated in the range of frequency
$10\,{\rm Hz}-1.5\,{\rm kHz}$ with a signal-to-noise ratio $\sim 3$ ($\sim 2$) for a pair of advanced LIGO
interferometers and if the dark-energy component is the cosmological constant (Chaplygin gas).

It is worth stressing that the sensitivity of the future third generation of detectors, as for example the Einstein
Telescope, could be high enough to increase the expected values of $({\rm S/N})$. For example, if we consider the 
ET-B sensitivity curve in triangular configuration the gain in relation to advanced LIGO would be $\sim 300-1000$.
Thus, instruments as ET could permit to explore the epoch when the first stars were formed in the
Universe at the end of the so-called `dark-ages'. In this way, the detection and characterization of a stochastic
background of GWs, over a large range in frequency, could be used as a tool for the study of the star formation up to
redshift $z\sim 20$.

Recently, \cite{mar1,mar2} also have analyzed stochastic backgrounds of gravitational waves but with a CSFR derived by \cite{tor07} which includes sources up to $z\sim 15$. Comparing the results of those papers with Fig. 1 of the present work, for the $\Lambda$-CDM case, we note that both CSFRs produce similar results up to $z\sim 3$. At higher redshifts ($z > 3$) the present study predicts more sources than TFS-CSFR (Tornatore, Ferrara \& Schneider CSFR). This happens because Pereira \& Miranda model (PM-CSFR) incorporates more baryons in stars than TFS-CSFR.

The preference for using PM-CSFR comes from the following points: (a) The dark energy component, with an equation of state dependent on time, modifies both the expansion factor $E(z)$ and the growth function (see equations \ref{sigmamz} and \ref{growth}). TFS-CSFR uses the GADGET code with the $\Lambda$-CDM model as the background cosmology. Thus, it would not be possible to study consistently the case Chaplygin gas as dark-energy fluid for the TFS-CSFR. This means that we would not have to compare their results with those obtained here for the case Chaplygin gas.

Note that one of our results was to show that the dark-energy component modifies the amplitude of the CSFR; (b) Recently \cite{pereira11} studied four different CSFRs (\citealp{pereira10}; \citealp{fardal}; \citealp{sh2003}; and \citealp{h4}) to derive the evolution of the comoving black hole mass density. Their results show that PM-CSFR has a good agreement with the quasar luminosity density up to redshift $\sim 6$. On the other hand, as PM-CSFR produces a high number of sources at $z > 3$ then there exist an important contribution of these objects, formed at high redshifts, to the backgrounds studied in the present paper. This is an intrinsic characteristic of the scenario used by \cite{pereira10} to derive the CSFR.

As a final point, different dark-energy scenarios could give different signatures on the background through the expansion
function $E(z)$. As the gravitational wave background could trace the behavior of the Universe up to redshift $\sim 20$,
then in principle it should be possible to infer if there is a temporal dependence of the dark-energy equation of state.
That is, if $\dot w(a) \neq 0$. In this way, not only binary systems at lower redshifts ($z< 2-3$) working as standard
sirens (e.g., \citealp{sathya10,zhao11}) but also stochastic backgrounds of gravitational waves could contribute for a
better comprehension of the physical nature of the dark energy. We observe that all viable dark-energy
fluids have similar behaviour up to $z\sim 1.5$ where the main observational data are available (e.g., SNIa and
baryon acoustic oscillations - BAO). This fact can be inferred from the CSFR (see Fig. 1) where for $z < 2$ all
models (cosmological constant and Chaplygin gas) have similar evolution. Thus, a way to identify if $\dot w(a) \neq 0$
would be observe the Universe at higher redshifts. In principle, stochastic backgrounds of gravitational waves
could be such observable.

\section{Acknowledgments}
ODM would like to thank the Brazilian Agency CNPq for partial financial support (grant 300713/2009-6). The author
would like to thank the referee, Tania Regimbau, for helpful comments that he feels considerably improved the paper.

\label{lastpage}

\end{document}